# Cosmology in two-dimensional dilaton gravity theories

M.A. AHMED

Department of Physics, Cornell University, Ithaca, New York 14853
&
Physics Department, Kuwait University, Kuwait.

Abstract

We study some aspects of classical & quantum cosmology in the context of two-dimensional dilaton gravity theories with matter being described by a perfect fluid. We derive the classical equations obeyed by the metric function & the dilaton field. An implicit solution for the metric function is obtained & used to discuss possible expansion & contraction scenarios for the universe. Quantization is carried out & finite norm exact solutions of the Wheeler-DeWitt equation are obtained.

PACS 04.60.Kz – Lower dimensional models; minisuperspace models.

# 1. Introduction

In quantum geometrodynamics one seeks to obtain a quantum theory of gravitation through quantization of the metrical structure of spacetime. It is difficult however to analyse quantum geometrodynamics for an arbitrary Riemannian spacetime. One then resorts to constructing models & the minisuperspace models of quantum cosmology provide a sutiable framework & have been extensively studied. In this approach all but a few of the degrees of freedom that describe the gravitational field & its sources are 'frozen out' [1]. Friedmann-Robertson-Walker (FRW) universes with varying matter content have been investigated in this manner [2].

Of special interest to us here is the study of the FRW model with matter described as a perfect fluid & quantization is performed using both the ADM method [3] & the method of superspace quantization [4,1]. The work of Lapchinski & Rubakov in this regard [5] was taken up, extended & applied in various directions by a number of authors [6] – [10]. Central to the work of Refs. [5] – [10] is the description of matter as a perfect fluid obeying the equation of state $p=k\rho$ where p is the pressure, $\rho$ is the total mass-energy density & k is a constant. The analyses were carried out using the canonical formulation of the generally relativistic hydrodynamics of a perfect fluid developed by Shutz [11], [12]. In this formulation velocity potentials are used to describe the motion of the fluid.

In the present work we use the aforementioned method to study the classical and quantum cosmology of two-dimensional dilaton gravity theories [13]. In an earlier work [14] we had studied this issue for the case of pressureless dust and by restricting ourselves to minisuperspace quantization. We were able to solve the classical equations for the scale factor and the dilaton field. We solved the Wheeler–DeWitt equation in the WKB approximation. Here we treat the case of nonzero pressure & employ the ADM quantization scheme. We set up the action for the model, obtain the super-Hamiltonian and derive the equations of motion for the metric function & the dilaton field. Although the solution for the metric function is given in an implicit form, we are still able to discuss various possible expansion & contraction scenarios for our model universe. We next quantize the model & cast the Wheeler-DeWitt equation in Schroedinger–like form that is amenable to analysis. An inner product is defined with respect to which the Hamiltonian is self-adjoint & boundary conditions on the wave function that ensure this crucial property are stated. The equation for the wave function of the universe is exactly solved & normalizable solutions are obtained. The material is organized as follows. In sect. 2 we set up the model & write down the action using the velocity potentials. The Hamiltonian is obtained & the equations of motion are stated. An implicit solution for the metric function is obtained & properties of the solution are discussed. In sec. 3 quantization of the model is carried out & the Schroedinger–like equation, arising from the DeWitt equation, satisfied by the wave function is written down & its solutions are derived. Sect. 4 is devoted to some concluding remarks.

## 2. Two-dimensional dilaton gravity

The two-dimensional spacetime is foliated by a one dimensional family of slices $\Sigma$, the slices being labeled by a time parameter t, t $\varepsilon$ $\Re$. Each t=const. slice $\Sigma$ has the topology of $\Re$ & is described by a coordinate parameter r. Furthermore each slice is spacelike & extends from left spatial infinity to right spatial infinity. The ADM form of the line element is

(2.1) $\quad ds^2 = -[N^2 - (N^r)^2 \wedge^2] dt^2 + 2\wedge^2 N^r dtdr + \wedge^2 dr^2$ .

Here $\wedge^2 dr^2$ is the induced spatial metric on $\Sigma$ and N & $N^r$ are the usual lapse & shift functions. Since $\wedge^2$ is the spatial metric $\wedge$ is nonzero. The action of two-dimensional dilaton gravity in the absence of matter reads [15]

(2.2) $\quad S_D = \int drdt \sqrt{-g} \, e^{-2\omega} \left[ R_o + 4(\nabla\omega)^2 + 4K^2 \right].$

In Eq.(2.2), $R_o$ denotes the Riemann curvature of the 2 metric $g_{\mu\nu}$, $\omega$ is the dilaton field & K is the cosmological constant. We shall describe matter by a perfect fluid with the action

(2.3) $$S_m = \int dr\, dt\, \sqrt{-g}\, p.$$

In ref. [14] we showed that only four velocity potentials are needed in 2 dimensions in contrast to the 4 - dimensional case of general relativity where six potentials are required [11]. In Shutz's notation these are $h, \phi, \theta$ & $S$ where h is the specific enthalpy and S is the specific entropy. In particular the 2-velocity is given by [14]

(2.4) $$U_\nu = h^{-1}\left(\phi_{,\nu} + \theta S_{,\nu}\right).$$

In ref. [14] we obtained the Hamiltonian formulation of the model. Restricting now ourselves to the homogeneous case where the metric functions, fields & momenta are functions of t only, we set $N^r = 0$ otherwise one would identify the particular one-dimensional vector $v = N^r$. We then find for the action $S = S_D + S_m$ the following expression.

(2.5) $$S = \int dt\left\{P_\wedge \dot{\wedge} + P_R \dot{R} + P_\phi (\dot{\phi} + \theta \dot{S}) - N H_o\right\},$$

where the dot denotes differentiation with respect to t & $H_o$ is the Hamiltonian density

(2.6) $$H_o = \frac{\wedge P_\wedge^2}{R^2} - \frac{P_\wedge P_R}{R} - K^2 R^2 \wedge + \wedge^{-k} P_\phi^{1+k}\, e^S.$$

The field R is related to the dilaton field w by $R = e^{-\omega}$ and $P_\wedge, P_R$ are the momenta canonically to $\wedge$ & R. Using $\theta = P_S/P_\phi$ [14] we can write

(2.7) $$S = \int dt\left(P_\wedge \dot{\wedge} + P_R \dot{R} + P_\phi \dot{\phi} + P_S \dot{S} - N H_o\right).$$

Let us now carry out the canonical transformation [10] that generalizes that of ref. [5]

(2.8) $$T = -P_S\, e^{-S}\, P_\phi^{-(k+1)},\quad P_T = P_\phi^{k+1}\, e^S,\quad \Phi = \phi + (k+1) P_S P_\phi^{-1},\quad P_\Phi = P_\phi.$$

We readily find that indeed the Poisson bracket (P.B) of T & $P_T$ is unity

(2.9) $$\{T, P_T\} = 1,$$

and that the action becomes

(2.10) $$S = \int dt \left( P_\Lambda \dot\Lambda + P_R \dot R + P_\Phi \dot\Phi + P_T \dot T - N\overline{H} \right),$$

where the super – Hamiltonian is given by

(2.11) $$\overline{H} = \frac{\Lambda P_\Lambda^2}{R^2} - \frac{P_\Lambda P_R}{R} - K^2 R^2 \Lambda + \Lambda^{-k} P_T .$$

The terms $P_\Phi$ & $\Phi$ do not occur in $\overline{H}$ & can be dropped from the action which now reads

(2.12) $$S = \int dt \left( P_\Lambda \dot\Lambda + P_R \dot R + P_T \dot T - N\overline{H} \right).$$

The phase space is generated by ($\Lambda$, R, T, $P_\Lambda$, $P_R$, $P_T$). The variable T is such that its P.B with $\overline{H}$ is

(2.13) $$\{T, \overline{H}\} = \Lambda^{-k} .$$

Since this P.B does not involve canonical momenta, it follow that T acts as a 'global time' using the terminology of [16].

The variational principle $\delta S = 0$ leads to the following classical equations of motion

(2.14) $$\dot\Lambda = N \left( \frac{2\Lambda P_\Lambda}{R^2} - \frac{P_R}{R} \right),$$

(2.15) $$\dot P_\Lambda = -N \left( \frac{P_\Lambda^2}{R^2} - K^2 R^2 \right),$$

(2.16) $$\dot R = -\frac{N P_\Lambda}{R},$$

(2.17) $$\dot P_R = -N \left( -\frac{2\Lambda P_\Lambda^2}{R^3} + \frac{P_\Lambda P_R}{R^2} - 2K^2 R \Lambda \right),$$

(2.18) $$\dot T = N \Lambda^{-k},$$

(2.19) $\dot{P}_T = 0$ .

These equations are supplemented by the super – Hamiltonian constraint

(2.20) $\overline{H} = \dfrac{\wedge P_\wedge^2}{R^2} - \dfrac{P_\wedge P_R}{R} - K^2 R^2 \wedge + \wedge^{-k} P_T = 0$ .

One can take the metric function $\wedge$ to be positive: $\wedge > 0$ [17]. Hence we can reparametrize $\wedge$ as follows

(2.21) $\wedge = e^{-\mu}$ ,

& recalling that $R = e^{-\omega}$ we then have that $P_\wedge = -e^\mu P_\mu$, $P_R = -e^\omega P_\omega$ where $P_\mu$ & $P_\omega$ are the momenta canonically conjugate to $\mu$ & $\omega$ respectively. We thus obtain that

(2.22) $\dot{\mu} = N e^{2\mu+\omega}(2P_\mu - P_\omega)$ ,

(2.23) $\dot{\omega} = -N e^{\mu+2\omega} P_\mu$ .

Next differentiating Eqs. (2.22) & (2.23) & using Eqs. (2.14) – (2.17) to calculate $\dot{P}_\mu$ & $\dot{P}_\omega$, leads to the following two equations

(2.24) $\ddot{\mu} - \dfrac{\dot{N}}{N}\dot{\mu} - (\dot{\mu} + 2\dot{\omega})\dot{\mu} = 0$ ,

and

(2.25) $\ddot{\omega} - \dot{\omega}^2 - \dfrac{\dot{\omega} \dot{N}}{N} - N^2 K^2 = 0$ .

In the gauge $t = T$ we have $N = \wedge^k$ and Eqs. (2.24) & (2.25) become

(2.26) $\ddot{\mu} - (1-k)\dot{\mu}^2 - 2\dot{\omega}\dot{\mu} = 0$ ,

and

(2.27) $\ddot{\omega} + k\dot{\omega}\dot{\mu} - \dot{\omega}^2 - K^2 e^{-2k\mu} = 0$ .

Next we differentiate Eq. (2.24) and substitute for the $\dot{\omega}$ & $\ddot{\omega}$ terms in the resulting equation using Eqs. (2.24) & (2.25). In this way we obtain the equation

(2.28) $\dot{\mu}\dddot{\mu} - \dfrac{3}{2}\ddot{\mu}^2 + k\dot{\mu}^2\ddot{\mu} - \dfrac{1}{2}(1-k^2)\dot{\mu}^4 - 2K^2\dot{\mu}^2 e^{-2k\mu} = 0$ .

Eq. (2.28) looks formidable but its analysis becomes manageable when we realize that it is an autonomous differential equation. This invites the substitution

(2.29) $\quad F(\mu) = \dot{\mu}(t)$,

which gives

(2.30) $\quad \ddot{\mu} = F \dfrac{dF}{d\mu}$,

(2.31) $\quad \dddot{\mu} = F \left( \dfrac{dF}{d\mu} \right)^2 + F^2 \dfrac{d^2 F}{d\mu^2}$.

Using Eqs. (2.29) – (2.31) in Eq. (2.28) leads to the equation

(2.32) $\quad F \dfrac{d^2 F}{d\mu^2} - \dfrac{1}{2} \left( \dfrac{dF}{d\mu} \right)^2 + kF \dfrac{dF}{d\mu} - \dfrac{1}{2}(1-k^2)F^2 - 2K^2 e^{-2k\mu} = 0$.

Next we make the further substitution

(2.33) $\quad F(\mu) = z(\mu) e^{-k\mu}$.

The equation satisfied by z(μ) then turns out to be

(2.34) $\quad z \dfrac{d^2 z}{d\mu^2} - \dfrac{1}{2} \left( \dfrac{dz}{d\mu} \right)^2 - \dfrac{1}{2} z^2 - 2K^2 = 0$.

We make one more substitution

(2.35) $\quad p(z) = \dfrac{dz}{d\mu}$,

leading to the equation

(2.36) $\quad zp \dfrac{dp}{dz} - \dfrac{1}{2} p^2 - \dfrac{1}{2} z^2 - 2K^2 = 0$.

Eq. (2.36) can be cast in the form

(2.37) $\quad z \dfrac{d}{dz}(p^2) - (p^2 + z^2) - 4K^2 = 0$,

which is a first order differential equation that is readily solved to yield

(2.38) $\quad p(z) = \pm (z^2 + Az - 4K^2)^{1/2}$,

where A is an arbitrary constant. Using Eq. (2.38) in Eq. (2.35) & integrating we obtain

$$(2.39) \qquad \int \frac{dz}{(z^2 + Az - 4K^2)^{1/2}} = \pm \mu + B \ .$$

where B is another arbitrary constant. One can show that no real solutions for $F = \dot{\mu}(t)$ are possible if $z(\mu)$ is constant. Hence $p(z) \neq 0$ and the quadratic expression under the square root in Eq. (2.38) is greater than zero to ensure real p(z). Performing the integration in Eq. (2.39) we obtain

$$(2.40) \qquad z + \frac{1}{2}A + (z^2 + Az - 4K^2)^{1/2} = De^{\pm \mu} \ ,$$

where D is another constant that can take positive or negative values. Next from Eqs. (2.29), (2.33) & (2.40) we obtain, after some algebra, 2 possible solutions for $\dot{\mu}(t)$

$$(2.41) \qquad \dot{\mu}(t) = \alpha_1 e^{(1-k)\mu} + \alpha_2 e^{-(1+k)\mu} + \alpha_3 e^{-k\mu} \ ,$$

and

$$(2.42) \qquad \dot{\mu}(t) = \alpha_1 e^{(1+k)\mu} + \alpha_2 e^{(1-k)\mu} + \alpha_3 e^{-k\mu} \ ,$$

where the constants $\alpha_j$ are given by

$$(2.43) \qquad \alpha_1 = \frac{1}{2}D, \ \alpha_2 = \frac{1}{2D}\left(4K^2 + \frac{1}{4}A^2\right), \ \alpha_3 = -\frac{1}{2}A \ .$$

Evidently the second solution given by Eq. (2.42) is obtained from the first one of Eq. (2.41) by the exchange $\alpha_1 \leftrightarrow \alpha_2$. It is also clear from Eq. (2.43) that $\alpha_1$ & $\alpha_2$ have the same sign. In this connection we also point out that the constant D cannot assume zero value since it is given by $D = \pm \exp B$ where B is the constant appearing in Eq. (2.39) & hence $\alpha_1$, $\alpha_2$ are both nonzero.

Let us now integrate Eq. (2.41) and express the result in terms of $\wedge$. We obtain for the case of stiff matter with k=1

(2.44)
$$\frac{1}{2\alpha_1} \ell n \left| \frac{\alpha_2 \wedge^2 + \alpha_3 \wedge + \alpha_1}{\alpha_2 \wedge_o^2 + \alpha_3 \wedge_o + \alpha_1} \right| - \frac{\alpha_3}{2\alpha_1^2 \beta} \times \left[ \tan^{-1}\left(\frac{\wedge^{-1} + \alpha_3/2\alpha_1}{\beta}\right) - \tan^{-1}\left(\frac{\wedge_o^{-1} + \alpha_3/2\alpha_1}{\beta}\right) \right] = t - t_o .$$

In Eq. (2.44) $\wedge$ is the metric spatial function at time t, $\wedge_o$ is its value at time $t_o$ and $\beta$ is a constant given by

(2.45) $$\beta = \frac{2K}{D} .$$

Eq. (2.44) constitutes an implicit solution of the metric function $\wedge$ (t).

We now ask whether, for some fixed $t_o$ & $\wedge_o$, Eq.(2.44) admits solutions $\wedge \to \infty$ or $\wedge \to o$ as $t \to \infty$ corresponding to expansion or contraction modes of the universe. For $\alpha_1 > o$ we clearly have an expansion mode with

(2.46) $$\wedge \approx e^{(t-t_o)/\alpha_1} ,$$

as $t \to \infty$. If $\alpha_1 < 0$ however, Eq. (2.44) cannot support an expansion scenario where $\wedge$ increases to large values because the first term on the L.H.S is negative and the second term is finite. However for $\alpha_1 < 0$ & $\alpha_3 < 0$, a contracting solution of Eq. (2.44) where $\wedge < \wedge_o$ for $t > t_o$ is certainly possible because both terms on the L.H.S. of the equation are positive. First of all let us point out that for $\alpha_1 < 0$, Eq. (2.44) cannot continue to hold for $t \to \infty$ because for this to happen the quadratic form in the argument of the logarithm must vanish. But one can readily show that the equation

(2.47) $$\alpha_2 \wedge^2 + \alpha_3 \wedge + \alpha_1 = 0 ,$$

does not have any real solutions & hence the argument of the logarithm cannot become zero. Now in principle starting from $t = t_o$ contraction down to $\wedge = 0$ can occur a later finite instant $t = T_c > t_o$ which from Eq. (2.44) is given by

(2.48) $$T_c = t_o + \frac{1}{2\alpha_1} \ell n \left| \frac{\alpha_1}{\alpha_2 \wedge_o^2 + \alpha_3 \wedge_o + \alpha_1} \right| - \frac{\alpha_3}{2\alpha_1^2 \beta} \left[ \frac{\pi}{2} - \tan^{-1}\left(\frac{\wedge_o^{-1} + \alpha_3/2\alpha_1}{\beta}\right) \right] .$$

With $\alpha_1$, $\alpha_3$ both negative the second term on the R.H.S of Eq. (2.48) is negative & the third is positive & contraction down to $\wedge = 0$ is possible at $T_c > t_o$ provided the sum of these two terms is positive. However classical analysis is not expected to hold at such small length scales & hence Eq. (2.48) cannot be expected to be strictly valid but only indicative. If $\alpha_1 < 0$ & $\alpha_3 > 0$, a contracting solution of Eq. (2.44) is possible if, for example, the following condition is satisfied.

$$(2.49) \quad \frac{1}{2\alpha_1} \ell n \left| \frac{\alpha_2 \wedge^2 + \alpha_3 \wedge + \alpha_1}{\alpha_2 \wedge_o^2 + \alpha_3 \wedge + \alpha_1} \right| \rangle \frac{\pi |\alpha_3|}{2\alpha_1^2 \beta} .$$

Contraction is thus not guaranteed in general for it depends on an intricate interplay of the parameters $\alpha_1$, $\alpha_2$, $\alpha_3$ & $\beta$ in such a way that a condition like Eq. (2.49), for example, is satisfied.

Next we turn to the so-called vacuum case corresponding to $k = -1$. Describing the vacuum as a perfect fluid with the barotropic equation of state $p = -\rho$, was employed in ref. [18] to determine the vacuum spectrum using thermodynamic and semiclassical considerations. In ref. [7] a similar description of the vacuum was used to quantize the de Sitter cosmological model. In our case here, integrating Eq. (2.42) with $k = -1$ leads to the result

$$(2.50) \quad -\frac{1}{2\alpha_2} \ell n \left| \frac{\alpha_2 \wedge^2 + \alpha_3 \wedge + \alpha_1}{\alpha_2 \wedge_o^2 + \alpha_3 \wedge_o + \alpha_1} \right| - \frac{\alpha_3}{2\alpha_1 \alpha_2 \beta} \times$$
$$\times \left[ \tan^{-1} \left( \frac{\wedge^{-1} + \alpha_3/2\alpha_1}{\beta} \right) - \tan^{-1} \left( \frac{\wedge_o^{-1} + \alpha_3/2\alpha_1}{\beta} \right) \right] = t - t_0 .$$

Eq. (2.50) admits expansion solutions for $\alpha_2 < 0$ with

$$(2.51) \quad \wedge \approx e^{|\alpha_2|(t-t_o)} ,$$

for $t \to \infty$. Contracting solutions with $\wedge < \wedge_o$ for $t > t_o$ again arise for $\alpha_2, \alpha_3, < 0$ with similar characteristics to the case of $\wedge$ solutions described by Eq. (2.44) as described above & there is no need to repeat such an analysis.

### 3. Quantization

Our super – Hamiltonian reads

(3.1) $$\overline{H} = R^{-2} \wedge P_\wedge^2 - R^{-1} P_\wedge P_R - K^2 R^2 \wedge + \wedge^{-k} P_T .$$

When constructing the quantum model, one has to address the issue of factor ordering. In the Wheeler– DeWitt quantization scheme one replaces the canonical momenta by operators according to

(3.2) $$P_\wedge \to -i \frac{\partial}{\partial \wedge} , P_R \to -i \frac{\partial}{\partial R} , P_T \to -i \frac{\partial}{\partial T} ,$$

and with the corresponding super-Hamiltonian operator $\hat{\overline{H}}$, the Wheeler-DeWitt equation reads

(3.3) $$\hat{\overline{H}} \Psi(\wedge, R, T) = 0 ,$$

where $\Psi(\wedge,R,T) = 0$ is the wave function of the universe. We shall restrict ourselves to the case k = 1 & adopt, for now an order of factors such that the factors R & $\wedge$ occur to the left of derivative operators. Eq. (3.3) can then be written in the Shroedinger form

(3.4) $$i \frac{\partial \Psi}{\partial T} = \hat{H} \Psi ,$$

where the Hamiltonian $\hat{H}$ is given by

(3.5) $$\hat{H} = -R^{-2} \wedge^2 \frac{\partial^2}{\partial \wedge^2} + \wedge R^{-1} \frac{\partial^2}{\partial \wedge \partial R} - K^2 R^2 \wedge^2 .$$

We want $\hat{H}$ to be self-adjoint with respect to a suitably defined inner product. To find the appropriate inner product we write $\hat{H}$ as

(3.6) $$\hat{H} = \wedge R^{-1} \hat{O} - K^2 R^2 \wedge^2 ,$$

where $\hat{O}$ is given by

(3.7) $$\hat{O} = -R^{-1} \wedge \frac{\partial^2}{\partial \wedge^2} + \frac{\partial^2}{\partial \wedge \partial R} .$$

Eq. (3.6) suggests the inner product be defined by

(3.8) $$(\Psi, \Phi) = \int_0^\infty dR \int_0^\infty d\wedge \wedge^{-1} R\, \Psi^*(\wedge, R, T) \Phi(\wedge, R, T) .$$

Clearly existence of the inner product requires the wave functions to vanish at $\wedge = 0$ & decay in a suitable manner as $\wedge \to \infty$. To achieve self-adjointnes for the Hamiltonian we replace the $\wedge \frac{\partial^2}{\partial \wedge^2}$ term in $\hat{O}$ by

(3.9) $$\frac{1}{2}\left( \wedge \frac{\partial^2}{\partial \wedge^2} + \frac{\partial^2}{\partial \wedge^2} \wedge \right).$$

This choice is of course not unique & we make it because of its simplicity. In addition the self-adjointness conditions & the existence of the inner product require the following asymptotic behaviour properties for the wave functions

(3.10) $$\lim_{\wedge \to \infty} (\wedge \Psi^* \Phi) = 0 ,$$

(3.11) $$\lim_{R \to \infty} (R^3 \Psi^* \Phi) = 0 .$$

As one can easily see, Eq. (3.10) implies that $\Psi$ should approach zero as $\wedge \to \infty$ at least as

(3.12) $$\Psi \sim \wedge^\alpha , \quad \alpha < -\tfrac{1}{2} .$$

Similarly Eq. (3.11) implies that $\Psi$ should approach zero as $R \to \infty$ at least as

(3.13) $$\Psi \sim R^\beta , \quad \beta < -\tfrac{3}{2} .$$

We also have the following conditions for the behaviour of the wave function at small values of the variables $\wedge$ & R. Namely as $\wedge \to 0$

(3.14) $$\Psi \sim \wedge^\sigma , \quad \sigma > 0 ,$$

& as R → 0

(3.15)    $\psi \sim R^{\delta}$ ,   $\delta > 0$ .

Turning now to Eq. (3.4) we shall seek solutions of the form

(3.16)    $\Psi(\wedge, R, T) = e^{-iET} \psi(\wedge, R),$

where E is real. Eq.(3.4) then becomes

(3.17)    $\left\{ \wedge R^{-1} \left[ -\frac{1}{2} R^{-1} \left( \wedge \frac{\partial^2}{\partial \wedge^2} + \frac{\partial^2}{\partial \wedge^2} \wedge \right) + \frac{\partial^2}{\partial R \partial \wedge} \right] - K^2 R^2 \wedge^2 \right\} \psi = E\psi .$

We re-write Eq.(3.17) as

(3.18)    $\left( -\wedge^2 \frac{\partial^2}{\partial \wedge^2} - \wedge \frac{\partial}{\partial \wedge} + \wedge R \frac{\partial^2}{\partial \wedge \partial R} - K^2 R^4 \wedge^2 \right) \psi = E R^2 \psi .$

One can easily show that this partial differential equation is hyperbolic at all points in the (∧,R) plane except along the lines ∧ = 0 & R = 0. It is well known that a hyperbolic partial differential equation can be cast into canonical form by introducing suitable new coordinates

(3.19)    $\xi = \xi(\wedge, R)$ ;   $\eta = \eta(\wedge, R).$

We find that the new variables $\xi$, $\eta$ must satisfy the following equations

(3.20)    $\frac{\xi_{\wedge}}{\xi_R} = 0$ ,

where $\xi_{\wedge} = \partial \xi / \partial \wedge$ etc., and

(3.21)    $\frac{\eta_{\wedge}}{\eta_R} = \frac{R}{\wedge} .$

Solving Eqs. (3.20) & (3.21) leads to

(3.22) $\quad \xi = R$,

(3.23) $\quad \eta = \ln(\wedge R)$.

In terms of $\xi$ & $\eta$, Eq. (3.18) takes the canonical form

(3.24) $\quad \dfrac{\partial^2 \psi}{\partial \xi \partial \eta} - \xi\left(K^2 e^{2\eta} + E\right)\psi = 0$,

where we continued to denote the wave function, expressed as a function of $\xi, \eta$ by the same symbol $\psi$. We seek a separable solution of Eq. (3.24) & write

(3.25) $\quad \psi(\xi,\eta) = X(\xi)Y(\eta)$.

Substitution in Eq.(3.24) then leads to the following equations for X & Y:

(3.26) $\quad \dfrac{dX}{d\xi} = \gamma \xi X$,

(3.27) $\quad \dfrac{dY}{d\eta} = \gamma^{-1}\left(E + K^2 e^{2\eta}\right)Y$,

where $\gamma$ is a separation constant. The solutions of Eqs. (3.26) & (3.27), expressed in terms of the original variables $\wedge$ & R, read

(3.28) $\quad X(R,\wedge) = \overline{X} \exp\left(\dfrac{1}{2}\gamma k^2\right)$,

&

(3.29) $\quad Y(R,\wedge) = \overline{Y}(R \wedge)^{\gamma^{-1}E} \exp\left(\dfrac{1}{2}\gamma^{-1}K^2 R^2 \wedge^2\right)$,

where $\overline{X}$ & $\overline{Y}$ are constants. The wave function $\psi(\wedge, R)$ is thus given by

(3.30) $\quad \psi(\wedge, R) = H(R \wedge)^{\gamma^{-1}E} \exp\left[\dfrac{1}{2}R^2\left(\gamma + \gamma^{-1}K^2 \wedge^2\right)\right]$,

where $H = \overline{X}\,\overline{Y}$. Now the solution given by Eq.(3.30) must satisfy the boundary conditions given in Eqs.(3.10a) – (3.11b). These require that $\gamma < 0$ & $\gamma^{-1}E > 1$. It follows then that $E < 0$. Since $\gamma$ is negative we write $\gamma = -s^2$ & rewrite Eq. (3.30) as

$$(3.31) \qquad \psi(\wedge, R) = H(R\wedge)^{-s^{-2}E} \exp\left[-\frac{1}{2}R^2\left(s^2 + s^{-2}K^2 \wedge^2\right)\right].$$

Eq. (3.31) is then our final result for the wave function of our model universe. We would like to point out solution given by Eq. (3.31) satisfies all the boundary conditions that we have stated earlier. As an illustration of the use of this result we calculate the expectation values $<\wedge>$ & $<R>$. These are defined by

$$(3.32) \qquad \langle \wedge \rangle = \frac{1}{\|\Psi\|^2} \int_o^\infty dR \int_o^\infty d\wedge R\, \Psi^*(R,\wedge,T)\Psi(R,\wedge,T),$$

$$(3.33) \qquad \langle R \rangle = \frac{1}{\|\Psi\|^2} \int_o^\infty dR \int_o^\infty d\wedge \wedge^{-1} R^2\, \Psi^*(R,\wedge,T)\Psi(R,\wedge,T),$$

where $\|\Psi\|^2 = (\Psi, \Psi)$. The integrations are readily done & we obtain the following results

$$(3.34) \qquad \|\Psi\|^2 = \frac{|H|^2}{4s^2}\left(\frac{K^2}{s^2}\right)^{-|E|/s^2} \Gamma\!\left(s^{-2}|E|\right),$$

$$(3.35) \qquad \langle \wedge \rangle = \frac{\sqrt{\pi}s^2 \Gamma\!\left(s^{-2}|E| + \frac{1}{2}\right)}{K\,\Gamma\!\left(s^{-2}|E|\right)},$$

$$(3.36) \qquad \langle R \rangle = \frac{1}{2}\left(\frac{\pi}{s^2}\right)^{1/2}.$$

Remarkably while $<\wedge>$ depends on E, s & K it turns out that the expectation value of the dilaton field $<R>$ is independent of E & K. Eqs. (3.35) & (3.36) lead to the following results for the behaviour of the expectation values at large & small values of $|E|$ & s. We have at fixed s:

(3.37) $$\lim_{|E|\to\infty} \langle\wedge\rangle = \frac{\sqrt{\pi}s^2}{K} ,$$

while at fixed $|E|$ we have

(3.38) $$\lim_{s\to\infty} \langle\wedge\rangle = \frac{\pi|E|}{K} .$$

Also we have at fixed $|E|$

(3.39) $$\lim_{s\to 0} \langle\wedge\rangle = 0 ,$$

& at fixed s we have

(3.40) $$\lim_{|E|\to 0} \langle\wedge\rangle = 0 .$$

It is also seen that <R> → 0 as s → ∞ but diverges as s → 0.

## 4. Conclusions

In this work we have investigated the two-dimensional gravity model with matter being described by a perfect fluid. This was done both classically & quantum mechanically & within the minisuperspace formulation. The velocity potential formalism of Shutz [11,12] was employed to describe the dynamics of the fluid. The classical equations of motion for the metric function & the dilaton field were derived. Solutions, implicit in nature, were obtained for the metric function for the case of stiff matter (k=1) & the vacuum case (k = −1) & used to study expansion & contraction scenarios for our model universe.

Next quantization of the model was carried out & a Schroedinger –like equation emanating from the Wheeler – DeWitt equation was written down. This partial differential equation turned out to be hyperbolic in nature & hence could be written in canonical form by an appropriate change of variables. In the canonical form the equation was solved exactly & a normalizable solution in closed form was obtained representing the wave function of the universe. This is to be contrasted with the wave functions obtained for FRW type models [6,8,9,10] or the de Sitter model [7] that were not normalizable. In these cases wave packets had to be constructed in order to arrive at normalizable wave functions & this entailed the rather arbitrary procedure of choosing weight functions in such a way as to be able to perform the integrations. No recourse to such procedure was necessary in our case. The expectation values of the metric function & the dilaton filed were calculated. A rather remarkable result was found in that the expectation value of the dilaton field <R> turned out to be independent of K & E, depending only on the separation variable $s^2$. It would be interesting to carry out an investigation of our model beyond the minisuperspace approximation especially with respect to the quantization aspect. We hope to return to this issue in the near future.

**Acknowledgements**

Some of this work was done while the author was spending part of his sabbatical leave at the Department of Physics of Cornell University. The author wishes to thank Professor Saul Teukolsky for his kind hospitality. The author also wishes to thank Kuwait University for financial support.